\documentclass[twocolumn, prb, aps, superscriptaddress, longbibliography,showpacs,amsmath,amssymb,floatfix
]{revtex4-2}

\usepackage{changes}
\usepackage{graphicx}
\usepackage{dcolumn}
\usepackage{bm}
\usepackage{graphicx}                                      
\usepackage{amssymb}

\usepackage{amsmath}
\usepackage{epsfig}
\usepackage{xcolor}
\usepackage{tabu}
\usepackage{mathtools}
\usepackage[colorlinks,linkcolor=blue,anchorcolor=blue,citecolor=blue,urlcolor=blue]{hyperref}
\usepackage{physics}
\usepackage{float}
\usepackage{diagbox}
\usepackage{inputenc}

\begin{document}

\title{Exact mobility edges in quasiperiodic network models with slowly varying potentials}
\author{Hai-Tao Hu}
\affiliation{Key Laboratory of Quantum Information, University of Science and Technology of China, Hefei 230026, China}
\affiliation{Anhui Province Key Laboratory of Quantum Network, University of Science and Technology of China, Hefei 230026, China}
\affiliation{Department of Physics, University of Science and Technology of China, Hefei, Anhui 230026, China}
\affiliation{Hefei National Laboratory, University of Science and Technology of China, Hefei 230088, China}
\author{Yang Chen}
\affiliation{Key Laboratory of Quantum Information, University of Science and Technology of China, Hefei 230026, China}
\affiliation{Anhui Province Key Laboratory of Quantum Network, University of Science and Technology of China, Hefei 230026, China}
\affiliation{Hefei National Laboratory, University of Science and Technology of China, Hefei 230088, China}
\author{Xiaoshui Lin}
\affiliation{Key Laboratory of Quantum Information, University of Science and Technology of China, Hefei 230026, China}
\affiliation{Hefei National Laboratory, University of Science and Technology of China, Hefei 230088, China}
\author{Ai-Min Guo}
\affiliation{Hunan Key Laboratory for Super-microstructure and Ultrafast Process, School of Physics, Central South University, Changsha 410083, China}
\author{Zijing Lin}
\email{zjlin@ustc.edu.cn}
\affiliation{Department of Physics, University of Science and Technology of China, Hefei, Anhui 230026, China}
\affiliation{Hefei National Laboratory, University of Science and Technology of China, Hefei 230088, China}
\author{Ming Gong}
\email{gongm@ustc.edu.cn}
\affiliation{Key Laboratory of Quantum Information, University of Science and Technology of China, Hefei 230026, China}
\affiliation{Anhui Province Key Laboratory of Quantum Network, University of Science and Technology of China, Hefei 230026, China}
\affiliation{Hefei National Laboratory, University of Science and Technology of China, Hefei 230088, China}
\affiliation{Synergetic Innovation Center of Quantum Information and Quantum Physics, University of Science and Technology of China, Hefei 230026, China}

\begin{abstract}
Quasiperiodic models are important physical platforms to explore Anderson transitions in low dimensional systems, yet the exact mobility edges (MEs) are generally hard to be determined analytically. To date, the MEs in only a few models can be determined exactly. In this manuscript, we propose a new class of network models characterized by quasiperiodic slowly varying potentials and the absence of hidden self-duality, and exactly determine their MEs. We take the mosaic models with slowly varying potentials as examples to illustrate this result and derive its MEs from the effective Hamiltonian. In this method, we can integrate out the periodic sites to obtain an effective Hamiltonian with energy-dependent potentials $g(E)V$ and effective eigenenergy $f(E)$, which directly yields the MEs at $f(E) = \pm(2t^\kappa \pm g(E)V)$, where $\kappa \in \mathbb{Z}^+$. With this idea in hand, we then generalize our method to more quasiperiodic network models, including those with much more complicated geometries and non-Hermitian features. Finally, we propose the realization of these models using optical waveguides and show that the Anderson transition can be observed even in small physical systems (with lattice sites about $L = 50 - 100$). Our results provide some key insights into the understanding and realization of exact MEs in experiments. 
\end{abstract}
\maketitle

\section{Introduction}
\label{sec-introduction}

Disorder is always an important subject of interest in condensed matter physics. The study of localization phenomena in disordered low-dimensional systems has been one of the central problems in condensed matter physics \cite{Schwartz2007Transport, Roati2007Anderson, Billy2008Anderson, Segev2013Anderson}. In one-dimensional (1D) systems, the scaling theory of localization indicates that any 
weak random potential leads all states to be localized \cite{Anderson1958Diffusion, Thouless1974Electrons, abrahams1979scaling, MacKinnon1981OneParameter,Sarker1981Scaling, evers2008anderson}. However, in models with correlated disorders, such as those involving quasiperiodic potentials \cite{aubry1980analyticity, Harper1955Single, Li2015Localization, Modak2015Many}, Anderson transitions occur at some finite critical strengths. Thus, 1D models with quasiperiodic potentials can exhibit much rich physics, including localized, extended, and critical phases, along with the associated mobility edges (MEs) \cite{Goblot2020criticality, Gao2024Probing, wang2025exact, zhou2025fundamental}. 
A pedagogical example of such a quasiperiodic system is the Aubry-Andr\'{e}-Harper (AAH) model \cite{aubry1980analyticity, cai2013topological, ganeshan2013topological, Wang2016Phase, Loughi2019Topological, Roy2021Reentrant}, in which the self-duality between real space and momentum space leads to the exact phase boundaries separating localized and extended states at some critical potential strength $V_{\rm c} = \pm 2t$ \cite{aubry1980analyticity}. Consequently, all states are extended when $|V| < |V_c|$, while they become localized for $|V| > |V_c|$, with all states at the critical points exhibit multifractality \cite{wilkinson1984critical, Siebesma1987Multifractal,tang1986scaling}. In Ref. \cite{hu2025hidden}, we show that this potential in the quasiperiodic network models can directly yield energy-dependent MEs, and in Ref. \cite{chen2025reentrant}, we investigate the Anderson transition in the quasiperiodic model using optical waveguides. Furthermore, the Anderson transitions in 1D models have been intensively explored in experiments with ultracold atoms \cite{Semeghini2015Measurement, Alexander2019Probing, Zeng2024Transition}, acoustic systems \cite{Ni2019Observation, Apigo2019Observation} and photonic systems \cite{Peng2020Localization, Wang2022Thouless, Alexey2023Anderson}. 

\begin{figure}
\includegraphics[width=0.48\textwidth]{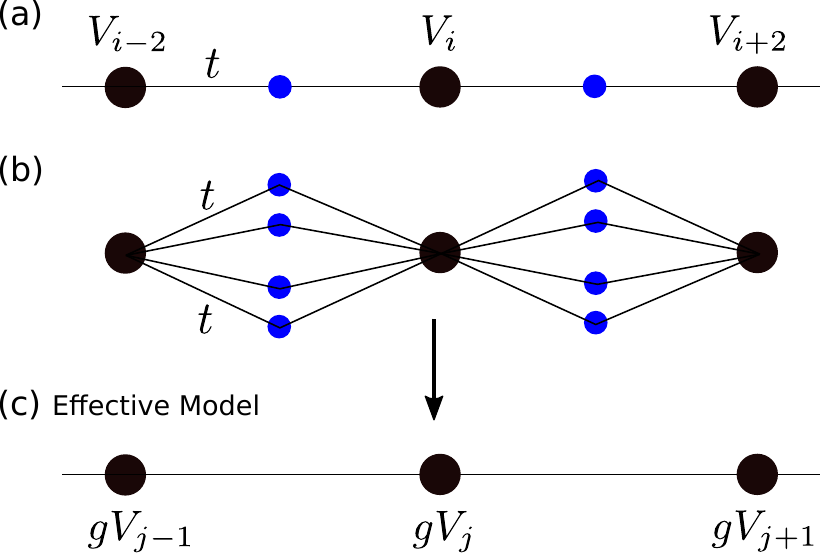}
\caption{(a) The 1D mosaic model with a slowly varying potential for $\kappa = 2$. (b) The network model when $N = 4$, which can be interpreted as a multipath mosaic model, where $N$ denotes the number of paths. More network models can be found in Ref. \cite{hu2025hidden}. (c) The effective quasiperiodic chain after integrating out the periodic sites. Here, $V_i$ represents the quasiperiodic slowly varying potential, for example $V \cos(\pi \alpha i^\nu + \phi)$, and $gV_j$ is the corresponding effective potential.}
\label{fig-fig1}
\end{figure}

The concept of self-duality has been generalized to other quasiperiodic systems, including AAH chains with exponentially decaying hopping terms \cite{biddle2010mobility, Liu2020Generalized}, nearest-neighbor tight-binding chains with other quasiperiodic potentials \cite{Ganeshan2015Mobility, Liu2020Generalized, Wang2021Duality}, and quasiperiodic mosaic AAH models \cite{Wang2020Quasiperiodic, liu2021mobility, wang2022topological, zhou2023mobility, dai2023multifractality, hu2025hidden}. In these models, the MEs can be directly determined from self-duality. However, there also exists a much broader class of quasiperiodic systems lacking self-duality \cite{biddle2011localization, deng2019quasicrystals, roy2021fraction, Fraxanet2022Localization, Yao2019Critical, Li2020Mobility, Lin2023critical, Lin2024Fate, li2025multifractal}, whose MEs may also be exactly solved using some other methods. For instance, several duality-breaking cases can be solved through the renormalization group method \cite{Goncalves2023Critical, Goncalves2023Renormalization}, and the quasiperiodic models with slowly varying potentials \begin{equation}
V_i = V{\rm cos}(\pi \alpha i^\nu),
\label{eq-ViSarma}
\end{equation}
can be exactly solved using semianalytical technique \cite{griniasty1988localization, thouless1988localization, das1988mobility, Das1990Localization, liu2018mobility, hu2025divergent}. In particular, Das Sarma {\it et al.} demonstrated that the 1D quasiperiodic models with this potential can support MEs at $E_c = \pm (2t - V)$ for $0 < \nu < 1$ and $V < 2$, while for $\nu > 1$, all states away from the exact band center are localized \cite{das1988mobility}. 

Recently, we demonstrated a method for determining the MEs of various quasiperiodic network models with self-dual potentials \cite{hu2025hidden}, which is constituted by periodic network sites and quasiperiodic sites. We show that although the original models lack self-duality during Fourier transformation to the momentum space, the effective models obtained by integrating out the periodic sites exhibit a hidden self-duality. This method naturally yields energy-dependent MEs. As compared to the above models, the quasiperiodic network models offer some significant flexibility, enabling us to design the MEs in more physical models. The major idea of this work is to extend the discussion of MEs from network models with self-dual potentials in Ref. \cite{hu2025hidden} to those with slowly varying potentials that lack self-duality. We theoretically obtain the exact MEs in a new class of 1D quasiperiodic network models with slowly varying potentials. We take the potential in Eq. (\ref{eq-ViSarma}) as an example to illustrate this idea. In our method, by integrating out the periodic sites, we obtain an effective tight-binding model, from which the exact MEs can be determined in terms of the energy-dependent potential $g(E)V$ and effective eigenenergy $f(E)$. Following the semianalytical method proposed by previous work \cite{das1988mobility}, we can directly obtain the MEs at $f(E) = \pm (2t\pm g(E)V)$. In this method, $g(E) = 0$ naturally yields the resonant states, which separate the extended states from the localized states. Finally, we propose the physical realization of these network models using optical waveguides and demonstrate that their associates Anderson transition in principle can be observed using some small physical systems, with lattice sites about $L = 50 -100$. In this way, we expect the physics models proposed in this work can be realized in the near future in experiments. 

\begin{figure}
\includegraphics[width=0.48\textwidth]{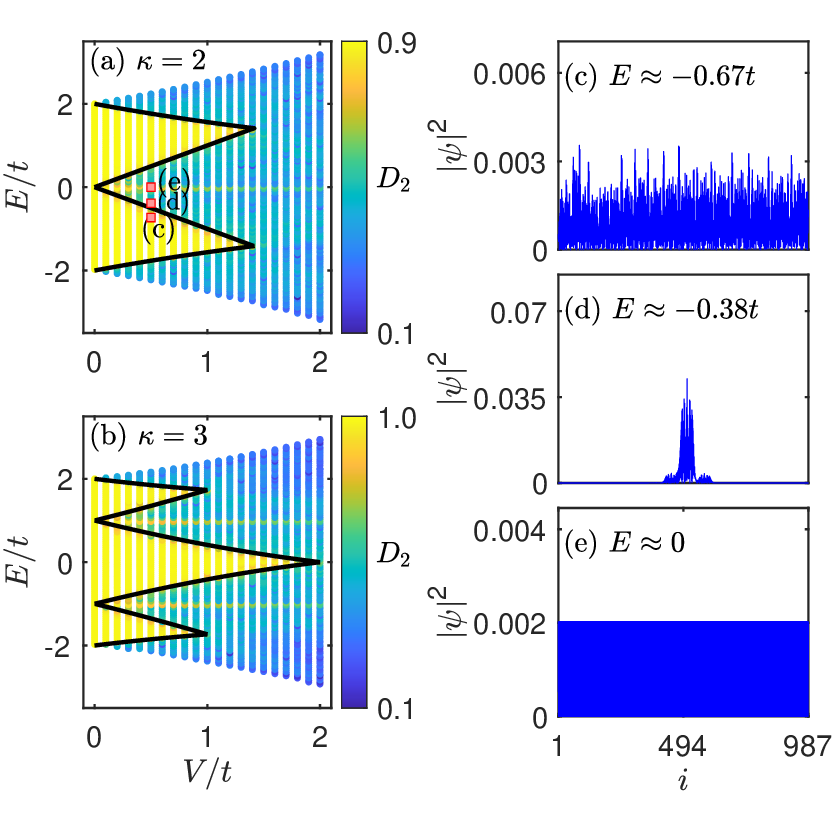}
\caption{Fractal dimension $D_2$ of different eigenstates as a function of energy $E$ and potential strength $V$ for mosaic models with slowly varying potentials when (a) $\kappa = 2$ and (b) $\kappa = 3$. The solid lines denote the MEs given by Eqs.~(\ref{eq-ME1}), (\ref{eq-ME2}), and (\ref{eq-ME3}), respectively. (c)-(e) Spatial distributions of three representative eigenstates with $E \approx -0.67t$ (extended), $E \approx -0.38t$ (localized), and $E = 0$ (resonant) as marked in Fig.~\ref{fig-fig2}(a). The other parameters are $\alpha = (\sqrt{5} - 1) / 2$, $\nu = 0.6$, $\phi = 0$ and $L = 987$.}
\label{fig-fig2}
\end{figure}

The rest of the manuscript is organized as follows. Section~\ref{sec-quasiperiodic mosaic model} introduces the Hamiltonian of mosaic models with slowly varying potentials and presents a method to solve their MEs. Next, in Sec.~\ref{sec-quasiperiodic network model}, we generalize our results to network models with slowly varying potentials and obtain their MEs. In Sec.~\ref{sec-experimental}, we propose to realize these models using optical waveguides and demonstrate that the transition from extended phase to localized phase may be observed even with small physical systems ($L = 50 - 100$), which is currently realizable in experiments \cite{chen2025reentrant}. Finally, we summarize and discuss these results in Sec.~\ref{sec-conclusions}.

\section{Mosaic models with slowly varying potentials}
\label{sec-quasiperiodic mosaic model}
\subsection{Hamiltonian and MEs}

The quasiperiodic mosaic models with slowly varying potentials, as illustrated in Fig.~\ref{fig-fig1}(a), can be described by the following Hamiltonian
\begin{align}
\mathcal{H} = -t \sum_{i} (c_{i + 1}^\dagger c_i + \mathrm{H.c.} ) + \sum_{i} V_i c_i^\dagger c_i,
\label{eq1}
\end{align}
where $c_i^\dagger$ ($c_i$) represents the creation (annihilation) operator at site $i$, and $t$ denotes the nearest-neighbor hopping coefficient. For simplicity, we set $t = 1$ as the energy unit. The on-site energy $V_i$ follows a mosaic pattern and is defined as \cite{Gong2021Exact}
\begin{align}
V_i &= \left\{
\begin{array}{lr}
V {\rm cos}(\pi \alpha i^\nu + \phi), \hspace{0.3cm} {\rm mod}(i,\kappa)=0, \\
0, \hspace{2.53cm} {\rm otherwise},
\end{array}
\right.
\label{eq2}
\end{align}where $\kappa$ is an integer, $V$ denotes the on-site potential strength, and $\phi$ is the phase offset. This potential is slightly different from that in Eq. (\ref{eq-ViSarma}). Without loss of generality, we set $\phi = 0$ and $\alpha = (\sqrt{5} - 1) / 2$. We have used the open boundary conditions to obtain the eigenvalues and eigenvectors \cite{hu2025hidden, Lin2023critical}. 

We can gain some insight into the physics of this model by considering some special limits, which have already been studied in literature. When $\nu = 1$ and $\kappa = 1$, the model reduces to the well-known AAH model, which exhibits a phase transition between extended and localized states. Specifically, all states are extended for $|V| < 2t$ and localized for $|V| > 2t$, with $V = \pm 2t$ being the self-dual points where all states are critical \cite{aubry1980analyticity}. For $\nu = 1$ and $\kappa \neq 1$, the model becomes the quasiperiodic mosaic AAH model, where the hidden self-duality ensures the presence of MEs, such as $E_c = \pm2t^2/V$ for the minimal case $\kappa = 2$, and $E_c = \pm \sqrt{t^2\pm2t^3/V}$ for $\kappa = 3$ \cite{Wang2020Quasiperiodic}.  When $0 < \nu < 1$ and $\kappa = 1$, the model corresponds to a quasiperiodic model with slowly varying potential, where extended states are located at the central region of the band ($|E| < 2t - V$) and localized states appear at the band edge ($|E| > 2t - V$). For $\nu > 1$ and $\kappa = 1$, all states away from the exact band center are localized \cite{das1988mobility}.

In this manuscript we focus on the condition that $0 < \nu < 1$ and $\kappa \neq 1$. We show that the MEs in this model can be exactly determined from the effective Hamiltonian, obtained by integrating out the periodic sites. For the minimal case of $\kappa = 2$, there are four exact energy-dependent MEs given by
\begin{align}
E_c = \pm V, \hspace{5mm} E_c = \pm \frac{\sqrt{V^2+16t^2} - V}{2},
\label{eq-ME1}
\end{align}
with a maximum potential strength of $V_\text{max} = \sqrt{2}t$. Additionally, there exists a resonant state at $E = 0$, where the wave function occupies only the sites with potential $V_i = 0$. Similarly, the MEs for $\kappa = 3$ are given by 
\begin{align}
E_c = \pm \frac{V+t-\sqrt{V^2-2Vt+9t^2}}{2},
\label{eq-ME2}
\end{align}
and
\begin{align}
E_c = \pm \frac{V-t\pm\sqrt{V^2+2Vt+9t^2}}{2},
\label{eq-ME3}
\end{align}
with $V_\text{max} = 2t$ for $E = 0$ and $V_\text{max} = t$ for $E = \pm \sqrt{3}t$. In this case, two resonant states exist at $E = \pm t$. These results may also be obtained using the transfer matrix method for the 1D Mosaic model in Fig. \ref{fig-fig1}(a) \cite{Gong2021Exact}.
In the following, we first present the numerical results in Sec.~\ref{sec-numerical}, followed by detailed derivations in Sec.~\ref{sec-shortcut}. 

\subsection{Numerical results}
\label{sec-numerical}

We can characterize the localization properties of the wave functions by calculating their inverse participation ratio (IPR) and normalized participation ratio (NPR). For the $n$-th normalized wave function $\ket{\psi^{n}} = \sum_i u_i^{n} c_i^\dagger \ket{0}$, the IPR and NPR are defined as 
\begin{align}
{\rm IPR^{n}} = \sum_{i = 1}^{L} |u_{i}^{n}|^4, \quad {\rm NPR}^n = \frac{1}{L} \frac{1}{\sum_{i = 1}^{L}|u_{i}^{n}|^4},
\end{align}
where $n$ represents the band index. For an extended eigenstate, with $u_i^n \sim 1/\sqrt{L}$, the IPR scales as ${\rm IPR}^{n} \propto L^{-1}$, which approaches zero in the thermodynamic limit. In this case, ${\rm NPR}^n \sim \mathcal{O}(1)$. Conversely, for a localized state, the IPR maintains a finite value as the system size increases, thus ${\rm IPR}^{n} \propto \mathcal{O}(1)$, thus 
${\rm NPR}^n \sim 1/L$. For a critical state, the IPR scales as ${\rm IPR}^{n} \propto L^{-D_2}$, where the fractal dimension $D_2 \in (0, 1)$. To distinguish the extended, localized, and critical states, we can define the fractal dimension as
\begin{align}
D_2^{n} = - {\ln {\rm IPR^{n}} \over \ln L}.
\end{align}
This number resembles the Hausdorff dimension defined in fractal geometries. This number has be used to characterize these different phases when and only when $L$ is large enough \cite{Lin2023critical}. 
 
\begin{figure}
\includegraphics[width=0.48\textwidth]{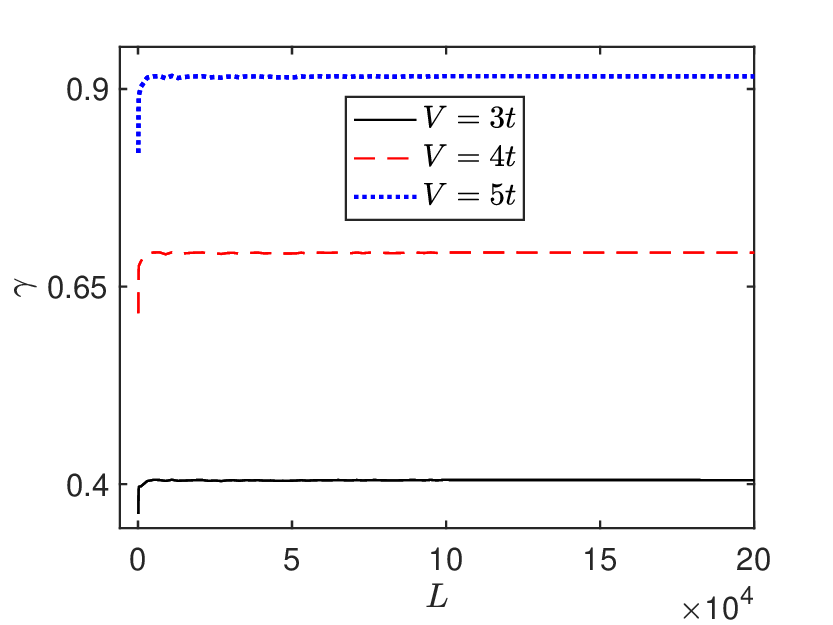}
\caption{Lyapunov exponent $\gamma$ as a function of chain length $L$ for the slowly varying models for various $V$ when $E = 0$. The Lyapunov exponent converges to a constant (in the localized phase) with the increasing of chain length $L$. The other parameters are the same as those in Fig. \ref{fig-fig2}.}
\label{fig-fig3}
\end{figure}

Figure \ref{fig-fig2}(a) plots the fractal dimension of the corresponding eigenstates as a function of the energy $E$ and potential strength $V$ for $\kappa = 2$. The black lines in the figure represent the MEs given by Eq.~(\ref{eq-ME1}). Consistent with the analytical results, the fractal dimension $D_2$ is approximately one for energies satisfying $V < |E| < (\sqrt{V^2+16t^2} - V)/2$ and zero for other energies. For any $\kappa$, the model exhibits $2 \kappa$ MEs, similar to the mosaic AAH model \cite{Wang2020Quasiperiodic}. The validity of the MEs is further confirmed by the spatial distributions of three typical eigenstates, marked by pink squares in Fig.~\ref{fig-fig2}(a). Figures~\ref{fig-fig2}(c)-\ref{fig-fig2}(e) display the spatial distribution of these eigenstates, corresponding to $E \approx -0.55t$ (extended), $-0.45t$ (localized), and $0$ (resonant) when $V = 0.5t$, respectively. These results are consistent with Eq.~(\ref{eq-ME1}), which predicts extended states for $0.5t < |E| < 1.76556t$. Additionally, we plot the fractal dimension of the corresponding eigenstates for $\kappa = 3$ in Fig.~\ref{fig-fig2}(b), which agrees well with the MEs given by Eqs.~(\ref{eq-ME2}) and (\ref{eq-ME3}).

\begin{figure}
\includegraphics[width=0.48\textwidth]{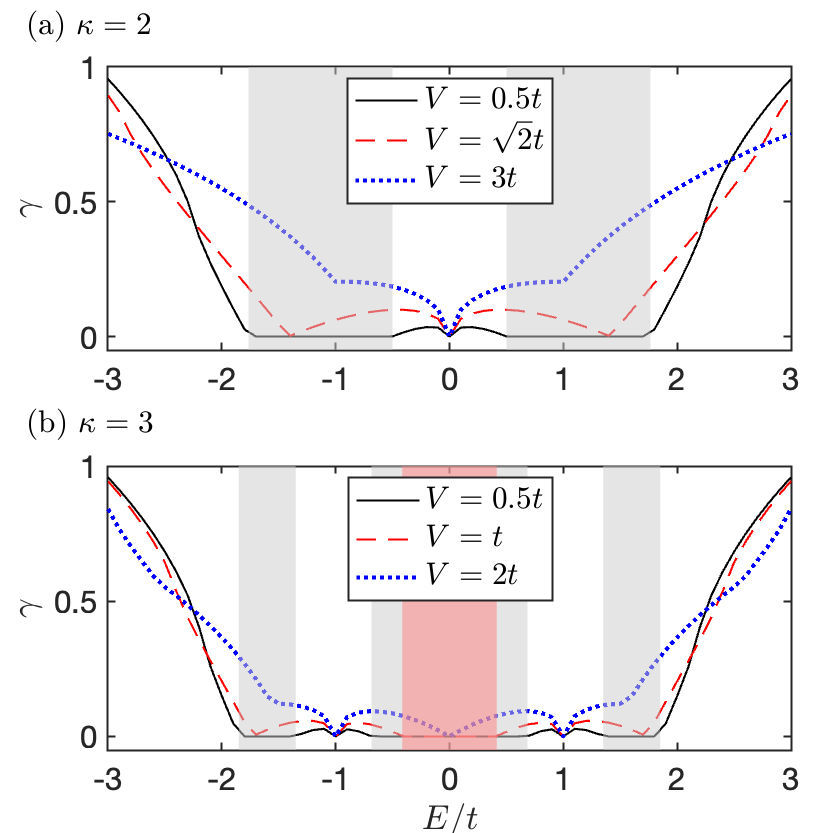}
\caption{Lyapunov exponent $\gamma$ as a function of energy $E$ for mosaic models with slowly varying potentials at various $V$ when (a) $\kappa = 2$ and (b) $\kappa = 3$. When the energy satisfies the conditions of Eqs.~(\ref{eq-ME1}), (\ref{eq-ME2}), and (\ref{eq-ME3}), $\gamma$ transitions from zero to a finite value, indicating the existence of MEs. The system size is $L = 832040$, and the other parameters are the same as those in Fig. \ref{fig-fig2}.}
\label{fig-fig4}
\end{figure}

The above analysis can give invaluable information about the localization properties of the wave functions. For the localized states, we expect that the wave functions follow 
\begin{align}
|\psi(x)| \sim e^{-\gamma (x-x_0)}.
\end{align}
Here, $\gamma = \gamma(E)$, termed as Lyapunov exponent, depends on the energy $E$, and $x_0$ is the corresponding localization center. Furthermore, we use the transfer matrix method to determine the Lyapunov exponent, in which $1/\gamma$ gives the localization length. Specifically, when the tight-binding model involves only nearest-neighbor hopping terms, i.e.,
\begin{align}
- \psi_{i - 1} - \psi_{i + 1} + V_i \psi_i = E \psi_i
, \label{eq8}
\end{align}
it can be transformed into the following form
\begin{align}
\begin{pmatrix}
\psi_{i+1} \\
\psi_i
\end{pmatrix}
=
T_i
\begin{pmatrix}
\psi_{i} \\
\psi_{i-1}
\end{pmatrix}
, \label{eq9}
\end{align}
where the transfer matrix $T_i$ is defined as
\begin{align}
T_i =
\begin{pmatrix}
{V_i - E} & -1 \\
1 & 0
\end{pmatrix}
. \label{eq10}
\end{align}
Then we can define a new matrix
\begin{align}
\mathbf{\Lambda} = \lim \limits_{L \rightarrow \infty} (\mathbf{T}_L^\dagger \mathbf{T}_L)^{1/(2L)}, \hspace{5mm} \mathbf{T}_L = \prod \limits_{i = 1}^L T_i.
 \label{eq11}
\end{align}
The Lyapunov exponents $\gamma$ is then obtained as 
\begin{align}
\gamma(E) = \text{min}_{i} ({\rm ln} \Lambda_i)
,
\end{align}
where $\Lambda_i$ represents the eigenvalues of the matrix $\mathbf{\Lambda}$. In this case, the smallest Lyapunov exponent corresponds to the condition with longest localization length by definition  $\gamma = 1/\xi$. Therefore, for the localized we expect $\gamma$ is finite; while for the extended phase, we expect that $\gamma L > 1$ (or $>> 1$), where $L$ is the chain length. We solve this matrix using the QR decomposition method. In Fig.~\ref{fig-fig3}, we plot the Lyapunov exponent $\gamma$ versus $L$ of the quasiperiodic slowly varying model for various $V$ when $E = 0$, showing that this value is convergent, which is necessary for a well-defined exponent. These results may also indicate ergodicity, which is necessary during the calculation of the above expression.  

\begin{figure}
\includegraphics[width=0.48\textwidth]{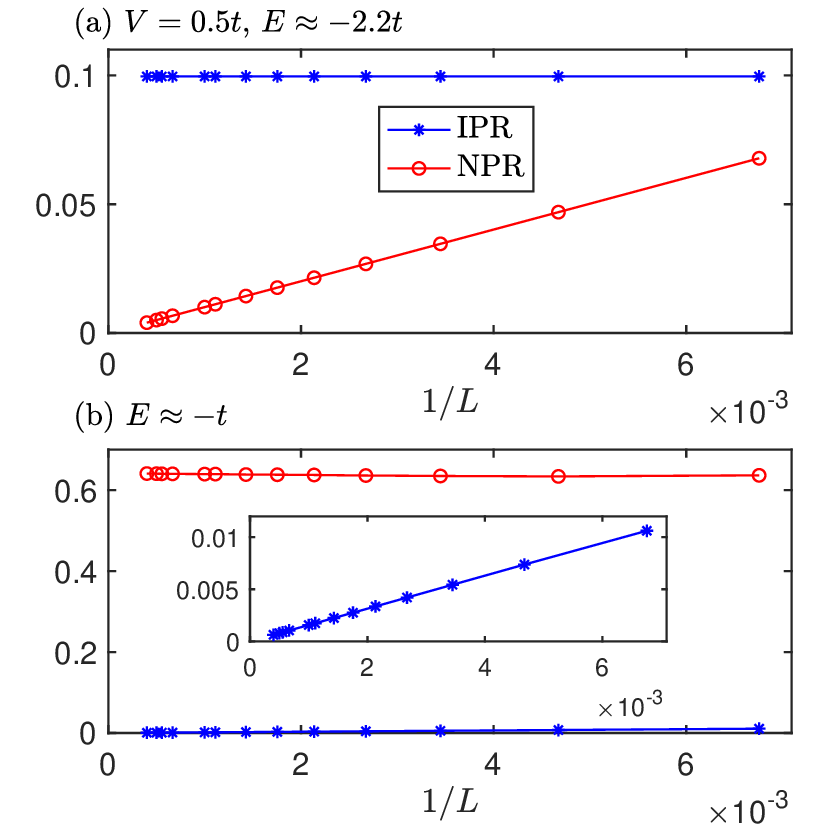}
\caption{Finite-size scaling of the IPR (blue asterisk) and NPR (red round) when $\kappa = 2$ and $V = 0.5t$ for (a) localized state $E \approx -2.2t$ and (b) extended state $E \approx -t$. The inset in (b) provides an enlarged view of the scaling of IPR for the extended state. The other parameters are the same as those in Fig. \ref{fig-fig2}.}
\label{fig-fig5} 
\end{figure}

To further investigate the localization properties of the quasiperiodic mosaic model with slowly varying potentials in the thermodynamic limit, we plot the Lyapunov exponent $\gamma$ as a function of energy $E$ for various $V$ in Fig.~\ref{fig-fig4}. As the energy varies, $\gamma$ changes dramatically from a finite value (indicating localized states) to 0 (indicating extended states), or vice versa, at certain energies. This abrupt jumping in $\gamma$ suggests the presence of MEs in the energy spectrum. When $\kappa = 2$, we observe that the Lyapunov exponent at the exact band center is vanishingly small ($\ll 1/L$) for different values of $V$ [see Fig.~\ref{fig-fig4}(a)], indicating a resonant state. The MEs are clearly observed around $E_c \approx \pm 1.8t$ and $\pm 0.5t$ for $V = 0.5t$, and around $E_c \approx \pm 1.4t$ for $V = \sqrt{2}t$, which closely align with the exact expression of MEs given by Eq.~(\ref{eq-ME1}). When $\kappa = 3$, we can also find the Lyapunov exponent $\gamma$ for states satisfying $E = \pm t$ is always zero, indicating two resonant states [see Fig.~\ref{fig-fig4}(b)]. Additionally, the regime with $\gamma = 0$ is consistent with Eqs.~(\ref{eq-ME2}) and (\ref{eq-ME3}).

In Fig.~\ref{fig-fig5}, we show the scaling behavior of the IPR and NPR for (a) the localized state $E \approx -2.2t$, and (b) the extended state $E \approx -t$ when $\kappa = 2$ and $V = 0.5t$. For the extended state, the NPR is independent of the system size, and the IPR decreases to zero following the law 
of $L^{-1}$, with $D_2 \rightarrow 1$, corresponding to an extended state. For the localized state, when $L \rightarrow \infty$, the NPR decreases to zero as $L^{-1}$  and the IPR remains constant. These results further demonstrate the existence of localization-delocalization transition in the model.

\subsection{Analytical solutions}
\label{sec-shortcut}

We now present a direct method to determine the MEs of the mosaic model with slowly varying potentials. By integrating out the periodic sites, we obtain an effective Hamiltonian with potential strength $g(E)V$ and eigenenergy $f(E)$, as described in Ref. \cite{hu2025hidden}. This effective model is expected to exhibit a localization-delocalization transition at $f(E)=\pm(2t^\kappa\pm g(E) V)$ ($\kappa \in \mathbb{Z}^+$), which gives the MEs. We take the case of $\kappa = 2$ as an example to illustrate this result, whose eigenvalue equations are given by
\begin{equation}
H {\bf u} = E {\bf u},
\end{equation}
where 
\begin{equation}
H = \begin{bmatrix}
\ddots & & & & & & & & \\
 & 0 & -t & & & & & & \\
 & -t & V_{i-2} & -t & & & & & \\
 & & -t & 0 & -t & & & & \\
 & & & -t & V_i & -t & & & \\
 & & & & -t & 0 & -t & & \\
 & & & & & -t & V_{i+2} & -t & \\
 & & & & & & -t & 0 & \\
 & & & & & & & & \ddots
\end{bmatrix},
\end{equation}
and 
\begin{equation}
{\bf u} = (
\cdots,
u_{i-2},
u_{i-1},
u_i,
u_{i+1},
u_{i+2},
\cdots)^T,
\end{equation}
with $T$ being the transpose. $H$ is a triangle matrix due to its hopping between nearest neighboring sites. From the equation for $u_{i-1}$ and $u_{i+1}$, where the potentials are zero, we can obtain
\begin{align}
u_{i-1} = \frac{-tu_{i-2}-tu_{i}}{E}, \quad u_{i+1} = \frac{-tu_{i}-tu_{i+2}}{E}
.
\end{align}
Substituting these expressions into the equation for $u_{i}$ yields
\begin{align}
(E^2-EV_{i}-2t^2) u_{i} & = t^2u_{i-2}+t^2u_{i+2}.
\end{align}
This equation involves only sites $u_i$ and $u_{i \pm 2}$, thus the periodic sites of $u_{i\pm 1}$ are integrated out. In this way, we can obtain an equivalent equation involving only quasiperiodic sites. Furthermore, let us re-define the operators using $u_i \rightarrow u_j$, with $j = [i/2]$, assuming that $i \in 2\mathbb{Z}$, we have  
\begin{align}
(E^2-EV_{j}-2t^2) u_{j} & = t^2u_{j-1}+t^2u_{j+1},
\end{align}
where $j \in \mathbb{Z}$ is continuous integer. The above equation can be re-organized as the following form 
\begin{align}
\lambda_j u_j  + t^2 u_{j-1} + t^2 u_{j+1}
= f(E) u_j,
\label{eq-eq6}
\end{align} 
where  
\begin{equation}
\lambda_j = g(E)V_{j}, \quad g(E) = E, \quad f(E) = E^2 - 2t^2
. 
\end{equation}
With effective potential strength $g(E)V$ and eigenenergy $f(E)$, we can directly obtian the MEs.

\begin{figure}
\includegraphics[width=0.48\textwidth]{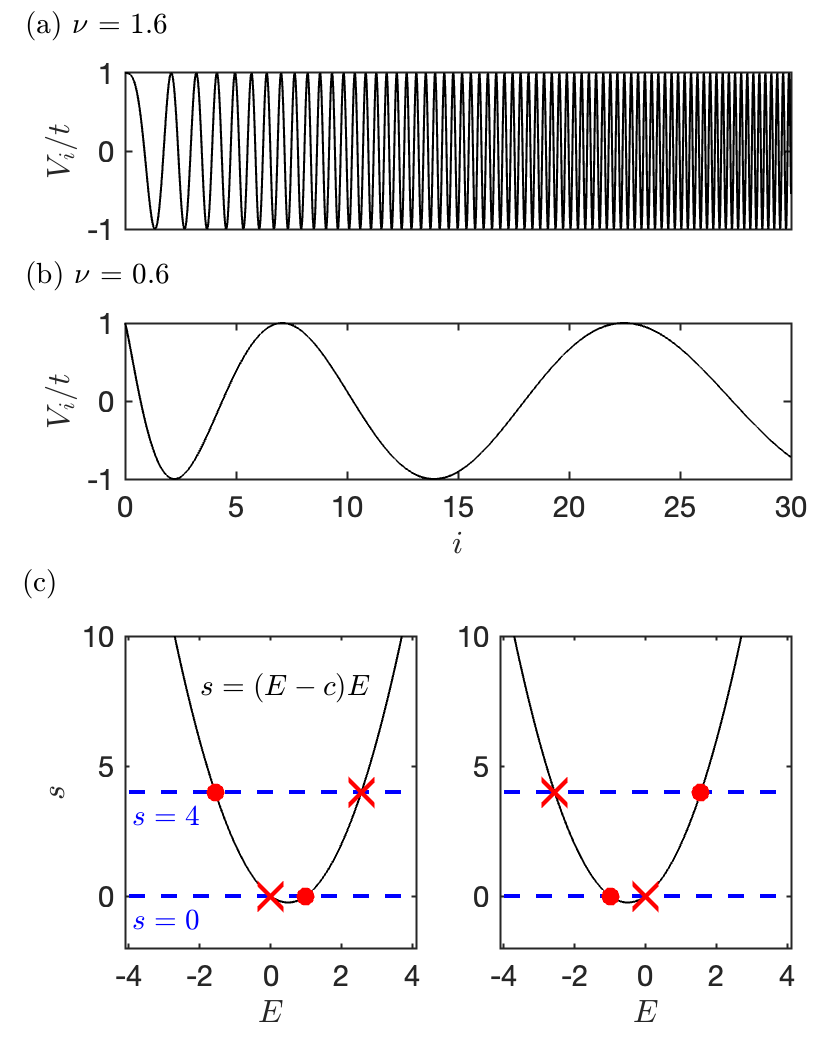}
\caption{Function $\cos(\pi \alpha j^{\nu})$ with (a) $\nu = 1.6$ and 
(b) $\nu = 0.6$. (c) Function $s = (E-c)E$ with $c > 0$ (left panel) and $c < 0$ (right panel), and their corresponding regime for $0 < s < 4t^2$, with $t=1$. The red dots denote the MEs.}
\label{fig-fig6}
\end{figure}

Next we provide a more detailed explanation of the above results using the method developed by Das Sarma {\it et al.} \cite{das1988mobility, Das1990Localization} in terms of "local constancy".
The effective potential $\lambda_j$ exhibits the property of becoming locally constant in the thermodynamic limit
\begin{align}
\lim \limits_{j \rightarrow \infty} |\frac{d\lambda_j}{dj}| = \lim \limits_{j \rightarrow \infty} -E V  \pi \alpha \nu \frac{{\rm sin}( \pi \alpha j^\nu + \phi)}{j^{1- \nu}} = 0
, \label{eq16}
\end{align}
due to $0 < \nu < 1$ [see Fig.~\ref{fig-fig6}(b)]. In contrast when $\nu > 1$, the oscillating of the potential will become more and more rapid, yielding fully localized states in the thermodynamic limit [see Fig.~\ref{fig-fig6}(a)].  

For the eigenstates of slowly varying models, one always expects the wave function with amplitude 
\begin{align}
u_j \sim z^j
, \label{eq17}
\end{align}
where $z$ is a complex quantity, thus $|z| =1$. The effective eigenvalue equation then becomes
\begin{align}
t^2 z^{j-1} + [(E - V_j)E - 2t^2] z^j + t^2 z^{j+1} = 0
, \label{eq18}
\end{align}
with 
\begin{align}
V_j = V {\rm cos}(\pi \alpha j^\nu + \phi)
, \label{eq19}
\end{align} 
which satisfies $|V_j| \leq V$. The crucial point is that when $V_j$ is locally constant for large $j$, in the vicinity of this regime, we can treat it to be a constant, that is 
\begin{align}
V_j \approx c \text{ locally independent of } j \text{ for large } j, 
\label{eq20}
\end{align}
with $-V \leq c \leq V$, yielding the following simplified equation
\begin{align}
t^2 z^{2} + [(E - c)E - 2t^2] z + t^2 = 0,
\label{eq21}
\end{align}
with the discriminant given by
\begin{align}
\Delta = [(E - c)E]^2 - 4t^2[(E - c)E],
\label{eq22}
\end{align}
and solution of $z$ given by 
\begin{equation}
z = {1\over 2t^2} (-E^2+c E +2t^2 \pm \sqrt{\Delta}).
\end{equation}
Based on this equation, Das Sarma {\it et al.} \cite{das1988mobility} pointed out that the real-complex transition of the above eigenvalue determine the MEs for extended-localized transition. Obviously, when $\Delta < 0$, the amplitude $z$ is complex, satisfying $|z| = 1$, which means that the state is extended; otherwise, it is a real value with $|z| \ne 1$, which gives the localized state. Then we can find that the condition for extended states at 
\begin{align}
0 < (E - c)E < 4t^2
. \label{eq23}
\end{align}
Obviously, it is a quadratic function of $E$, yielding the curve in Fig. \ref{fig-fig6}(c). When $c = V$, we find
\begin{equation}
E = {1\over 2} (V \pm \sqrt{16t^2 + V^2}), \quad E =0, \quad E=V,
\label{eq-eq32}
\end{equation}
and when $c = -V$, we have 
\begin{equation}
E = {1\over 2} (-V \pm \sqrt{16t^2 + V^2}), \quad E =0, \quad E=-V.
\label{eq-eq33}
\end{equation}
These solutions correspond the MEs given by 
Eq.~(\ref{eq-ME1}), resonant states $E = 0$, and energy spectrum boundaries. 

\begin{figure*}    \includegraphics[width=0.98\textwidth]{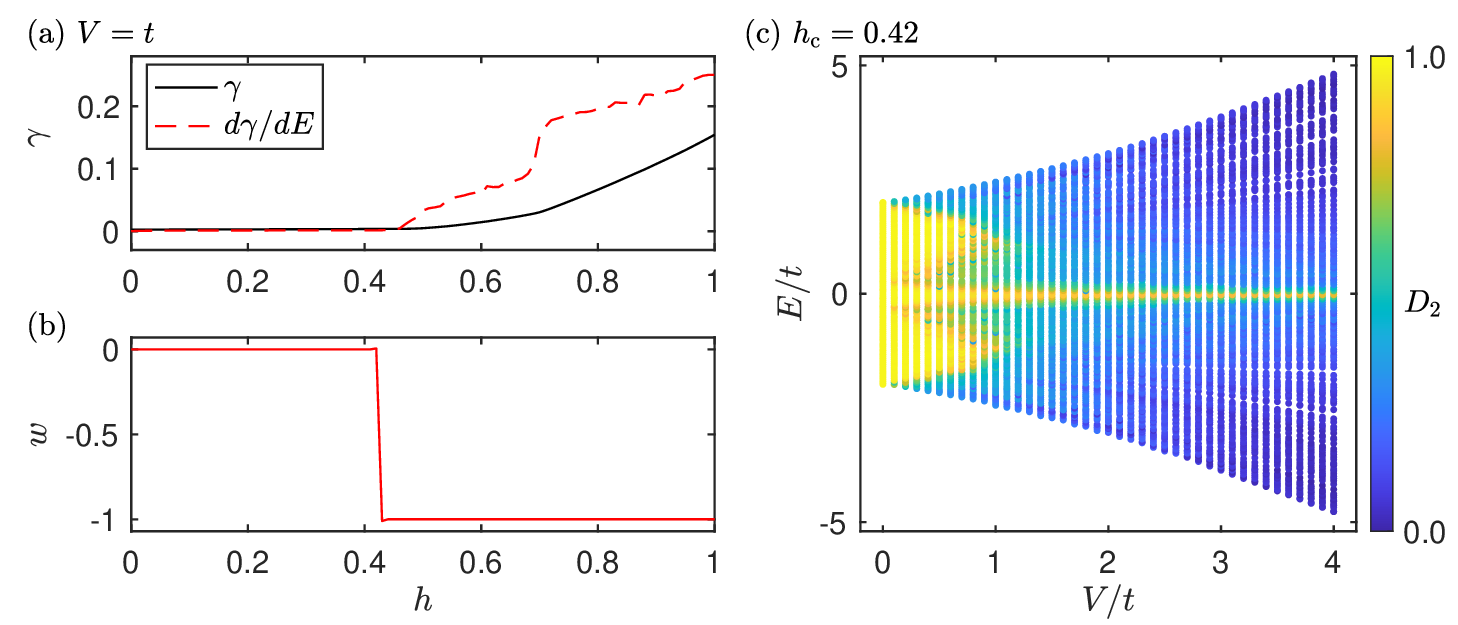}
\caption{(a) Lyapunov exponent $\gamma$ and its derivative $d\gamma/dE$ and (b) winding number $w$ versus phase $h$ in the quasiperiodic potential $V_i = \cos (\pi \alpha i^\nu + ih)$. (c) $D_2$ of the eigenstates as functions of $E$ and $V$ when $h = h_{\rm c} = 0.42t$. Here we can see that the localization-delocalization transition points in (a) and (c) are also topological transition points in (b). Here we choose quasiperiodic mosaic slowly varying models with $\kappa = 2$ as an example.}
\label{figR1}
\end{figure*}

The above results have an immediate corollary, that is, the phase boundaries are independent of the phase carried in the potential, assuming $V_j = V \cos(\pi \alpha j^{\nu} + \phi)$. The reason is that while this phase will influence the potential at each site, it will not influence the maximal and minimal value of the potential. Meanwhile, the parameters of $\alpha$ is also not important, nor it is not necessary to be an irrational number. We find that $\alpha$ and $\phi$ will never appear in the expression of MEs. These are also verified by our numerical simulation, which are not shown. However, if the phase $\phi$ becomes complex, \textcolor{black}{$\phi = \theta+ih$, the physics may be totally different \cite{Padhan2024Complete, Loughi2019Topological}. To characterize the topological property induced by complex phase, we can define a winding number as
\begin{align}
w(h) = \lim_{L \rightarrow \infty} \frac{1}{2\pi i} \int_0^{2\pi} d\theta \frac{\partial}{\partial \theta}\text{log det}[H(\frac{\theta}{L},h)-\epsilon],
\end{align}
where we fix the base energy $\epsilon = 0$. In Fig.~\ref{figR1}(b), we numerically calculate the winding number $w$ as a function of $h$ for the quasiperiodic mosaic slowly varying model with $\kappa=2$ and $V=t$, revealing a topological transition point at $h_{\rm c} \approx 0.42$. To verify whether the topological transition also coincides with a localization-delocalization transition, we present the corresponding fractal dimension $D_2$ as a function of $E$ and $V$ when $h = h_{\rm c}$ in Fig.~\ref{figR1}(c). It is observed that as $V$ increases to $t$, all states except for the resonant state at $E=0$ become localized, indicating a transition from critical phase to localized phase. In Fig.~\ref{figR1}(a), we plot the Lyapunov exponent and its derivative for $V=t$ and $E=t$, which agrees well with the topological transition points $h_{\rm c}\approx0.42$.}

\begin{figure}
\includegraphics[width=0.48\textwidth]{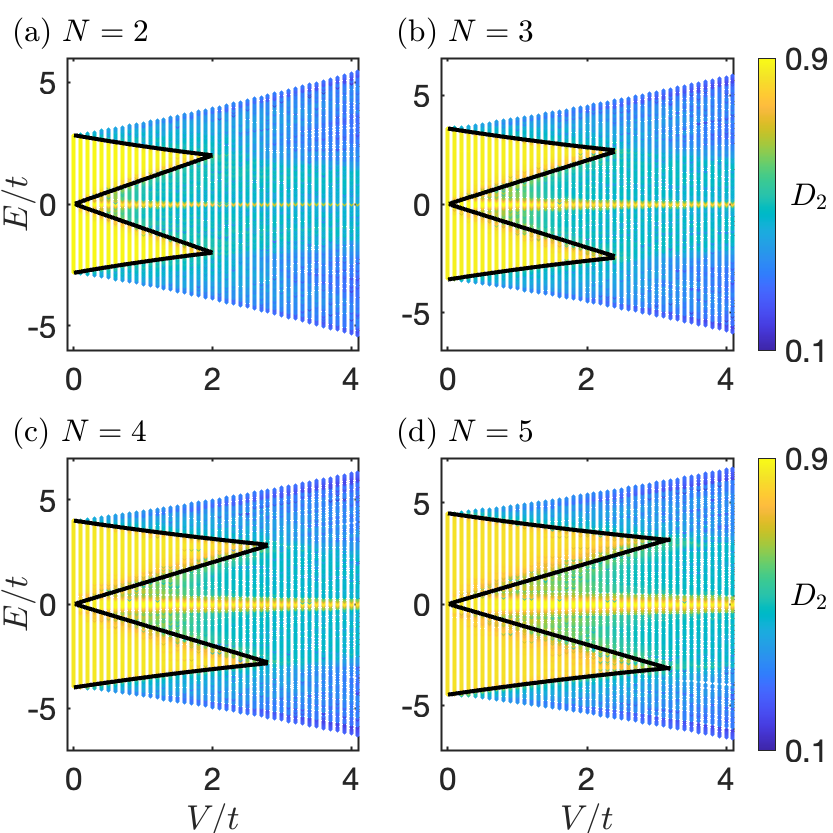}
\caption{Fractal dimension $D_2$ of different eigenstates as a function of energy $E$ and potential strength $V$ for multipath chain with slowly varying potentials when (a) $N = 2$, (b) $N = 3$, (c) $N = 4$, and (d) $N = 5$. The solid lines denote the MEs described by $f(E) = \pm(2t^2 \pm g(E)V)$ with function $f(E)$ and $g(E)$ given in Eq.~(\ref{eq-fg}). The other parameters are the same as Fig.~\ref{fig-fig2}.}
\label{fig-fig7}
\end{figure}

\section{Network models with slowly varying potentials}
\label{sec-quasiperiodic network model}

\subsection{Hermitian models}

The above method with effective Hamiltonian can be used to determine the MEs in more generalized quasiperiodic network models with slowly varying potentials \cite{hu2025hidden}. Consider a concrete case that can be described by the Hamiltonian
\begin{align}
\mathcal{H} & = \sum_{i = 1}^L (V_i c_i^\dagger c_i + \sum_{k=1}^N U_k d_{i,k}^\dagger d_{i,k})  - t (\sum_{i=1}^L \sum_{k =1}^N c_{i}^\dagger d_{i,k} \nonumber \\
& + \sum_{i=1}^{L-1} \sum_{k =1}^N c_{i+1}^\dagger d_{i,k} + \mathrm{H.c.}),
\label{eq-Hami2}
\end{align}
where $d_{i,k}^\dagger$ ($d_{i,k}$) is the creation (annihilation) operator at the $k$-th site of cell $i$, $V_i$ denotes the quasiperiodic slowly varying potentials defined as $V_i = V \cos(\pi \alpha i^\nu)$, and $U_k$ represents constant potentials. Here we set $U_k = 0$ for simplicity, and the results for nonzero cases are similar. Fig.~\ref{fig-fig1}(b) illustrates the multipath quasiperiodic network model with $N = 4$. For such a multipath quasiperiodic model, the MEs can also be obtained through the effective Hamiltonian. To see that, let us define the $n$-th eigenvector as $\ket{\psi^{n}} = \sum_i (u_i^{n} c_i^\dagger + \sum_k v_{i, k}^n d_{i, k}^\dagger) \ket{0}$, then from the following operator equation
\begin{equation}
E d_{i-1,k} = [d_{i-1,k}, H] = -t(c_{i-1} + c_{i}),
\end{equation}
the eigenvalue equation for the periodic sites, applying the eigenvector into the above equation, which is equivalent to the replacement 
\begin{equation}
d_{i,k} \rightarrow v_{i,k}, \quad c_i \rightarrow u_i,
\label{eq-replacement}
\end{equation}
will yields
\begin{align}
v_{i-1,k} = \frac{-tu_{i-1}-tu_{i}}{E}, \quad v_{i,k} = \frac{-tu_i-tu_{i+1}}{E}.
\end{align}
Substituting these results into the eigenvalue equation for $u_i$ from the following operator equation
\begin{equation}
Ec_{i} = [c_i, H] = V_i c_i -t\sum_{k}(d_{i,k} + d_{i-i, k}),
\end{equation}
will yields
\begin{align}
(E-V_i)u_{i} = \frac{t^2N(u_{i-1}+2u_{i}+u_{i+1})}{E}.
\label{eq-secIAmodel}
\end{align}
Here, the coefficient $N$ comes from the $N$ periodic sites; see $N=4$ in Fig. \ref{fig-fig1}(b).
Further, let us rewrite $u_i$ as $u_j$ to distinguish the index between the effective chain and original chain, we can obtain the effective Hamiltonian
\begin{align}
\sum_{j} g(E) V_j u_j + t^2 u_{j-1} + t^2 u_{j+1} = f(E) u_{j},
\end{align}
with 
\begin{align}
g(E) = \frac{E}{N}, \hspace{5mm} f(E) = -2t^2+\frac{E^2}{N},
\label{eq-fg}
\end{align}
which leads to the MEs at $f(E) = \pm [2t^2 \pm g(E)V]$.

Next, we verify the above results using numerical method. In Fig.~\ref{fig-fig7}, the fractal dimension $D_2$ of the eigenstates is plotted as a function of energy and potential strength for various $N$. The black lines represent the MEs described by $f(E) = \pm(2t^2 \pm g(E)V)$, with the corresponding $f(E)$ and $g(E)$ given by Eq.~(\ref{eq-fg}). This equation have four solutions, thus four MEs, which is independent of $N$. Additionally, there always exists a state at $E = 0$ as a resonant state, which remains extended regardless of $V$. This state is given by $g(E) = 0$. 

\begin{figure}
\includegraphics[width=0.48\textwidth]{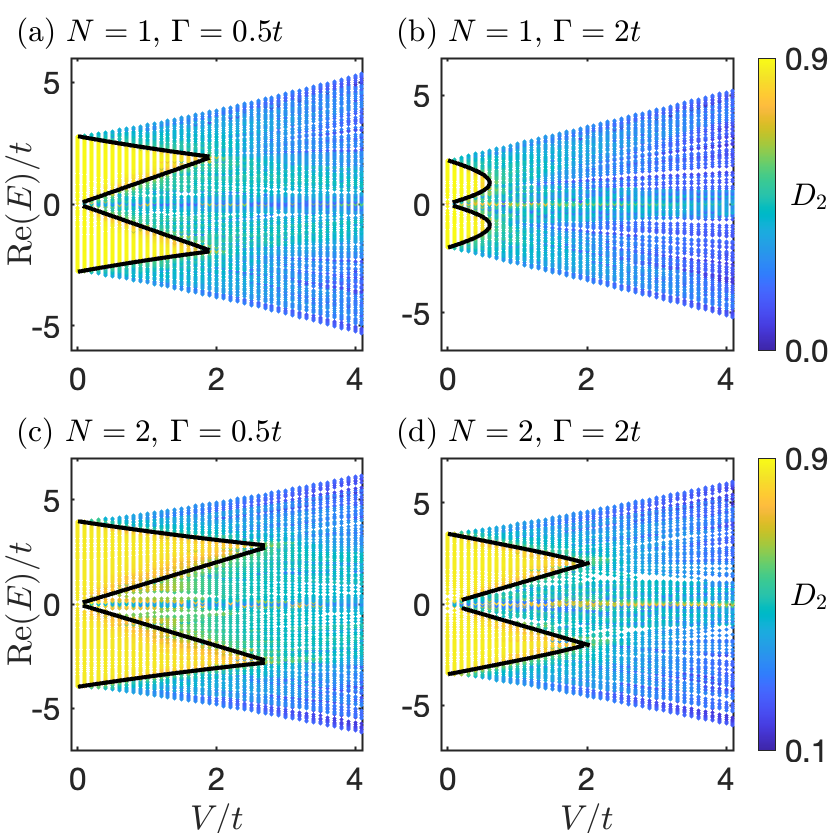}
\caption{Fractal dimension $D_2$ of different eigenstates as a function of energy $E$ and potential strength $V$ for PT-symmetric multipath chain with slowly varying potentials when (a) $N = 1$, $\Gamma = 0.5t$, (b) $N = 1$, $\Gamma = 2t$, (c) $N = 2$, $\Gamma = 0.5t$, and (d) $N = 2$, $\Gamma = 2t$. The solid lines denote the MEs described by $f(E) = \pm(2t^2\pm g(E)V)$ with function $f(E)$ and $g(E)$ given in Eq.~(\ref{eq-fg2}). The other parameters are the same as Fig.~\ref{fig-fig2}.}
\label{fig-fig8}
\end{figure}

\subsection{Non-Hermitian models}
The results for the non-Hermitian quasiperiodic network models with slowly varying potentials can be obtained in the same way. Here, we consider a special case with parity-time (PT) symmetry, which can be described as
\begin{align}
\mathcal{H} & = \sum_{i = 1}^L (V_i c_i^\dagger c_i + \sum_{k=1}^{N} i \Gamma d_{i,2k}^\dagger d_{i,2k} - i \Gamma d_{i,2k-1}^\dagger d_{i,2k-1}) \nonumber \\
& - t (\sum_{i=1}^L \sum_{k' =1}^{2N} c_{i}^\dagger d_{i,k'}
+ \sum_{i=1}^{L-1} \sum_{k' =1}^{2N} c_{i+1}^\dagger d_{i,k'} + \mathrm{H.c.}).
\end{align}
We still assume $\ket{\psi^{n}} = \sum_i (u_i^{n} c_i^\dagger + \sum_k v_{i, k}^n d_{i, k}^\dagger) \ket{0}$. 
The PT symmetry is realized by balancing the gain and loss in the periodic sites since the even sites $d_{i, 2k}$ and odd sites $d_{i, 2k-1}$ have opposite imaginary potentials. The eigenvalue equations for the periodic sites from the following operator equation
\begin{equation}
E d_{i-1,k} = [d_{i-1,k}, H] = -t(c_{i-1} + c_{i}) + U_{k} d_{i-1, k},
\end{equation}
and the replacement rule in Eq. \ref{eq-replacement} will yields
\begin{align}
v_{i,2k} = \frac{-tu_{i}-tu_{i+1}}{E- i\Gamma}, \quad v_{i,2k-1} = \frac{-tu_i-tu_{i+1}}{E+i\Gamma},
\end{align}
Substituting these results into the eigenvalue equation of $u_i$ will yield
\begin{align}
(E - V_i) u_{i} = \frac{2N Et^2(u_{i-1}+2u_{i}+u_{i+1})}{E^2+\Gamma^2}.
\end{align}
This equation is essentially the same as Eq. \ref{eq-secIAmodel}. Obviously, when the complex potentials are not opposite, this equation will become complex, yielding an possible route for non-Hermitian physics \cite{Padhan2024Complete, Loughi2019Topological}. 

\begin{figure*}
\includegraphics[width=0.98\textwidth]{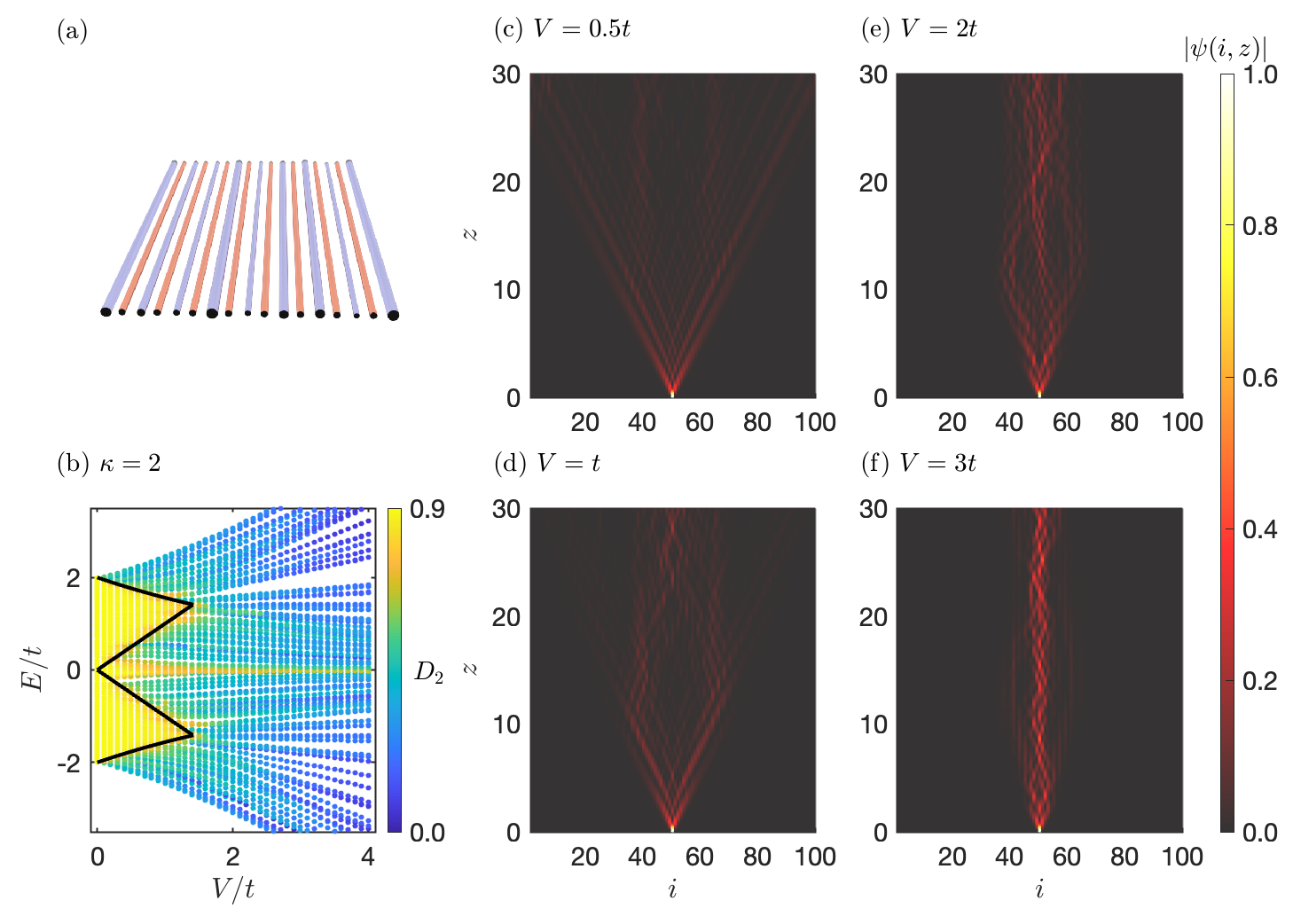}
\caption{(a) An illustration of the experimental setup based on optical waveguides for quasiperiodic mosaic model with $\kappa = 2$. Here, the blue waveguides represent the  quasiperiodic sites with slowly varying potentials, while the red ones represent the periodic sites with zero potential. (b) Fractal dimension $D_2$ of different eigenstates as a function of energy $E$ and potential strength $V$ for $L = 100$. The solid lines denote the MEs described by Eq.~(\ref{eq-ME1}). Evolution of the wave function $|\psi(i,z)|$ for the initial state $\psi(i,0) = \delta_{i,50}$ when (c) $V = 0.5t$, (d) $V = t$, (e) $V = 2t$, and (f) $V = 3t$.}
\label{fig-fig9}
\end{figure*}

Similarly, let us rewrite $u_i$ using $u_j$ and we can obtain the following effective Hamiltonian
\begin{align}
\sum_{j} g(E) V_j u_j + t^2 u_{j-1} + t^2 u_{j+1} = f(E) u_{j},
\end{align}
with 
\begin{align}
g(E) = \frac{(E^2+\Gamma^2)}{2N E}, \hspace{5mm} f(E) = -2t^2+\frac{E^2+\Gamma^2}{2N}.
\label{eq-fg2}
\end{align}
This solution will yields  MEs at $f(E) = \pm (2t^2 \pm g(E)V)$, which is numerically verified for various $N$ and $\Gamma$ in Fig.~\ref{fig-fig8}. The results for the other non-Hermitian cases can be derived in the same way.

\section{Experimental Proposal} 
\label{sec-experimental}

Finally, we discuss the possible way to realize these network models. The realization of quasiperiodic network models with various periodic network sites is challenging in cold-atom systems \cite{Christian2017Quamtum, Browaeys2020Many, Schafer2020Tools}, however, it can be easily realized using the coupled single-mode optical waveguides \cite{Christodoulides2003Discretizing, Lahini2009Observation, Kraus2012Topological, Verbin2013Observation, Tang2018Experimental}. By controlling the distance between neighboring waveguides and tuning their refraction index, we can modulate the hopping strength and on-site potential \cite{Alexander2005Discrete, Szameit2007Control, Chen2021Tight}, enabling the realization of various quasiperiodic network models with slowly varying potentials. To observe the localized-delocalized transition in such systems, we can inject light into a single waveguide and measure the distribution of light intensity after it propagates through a certain distance in the waveguides array [see Fig.~\ref{fig-fig9}(a)]. The propagation of light along the waveguide can be described by tight-binding equation
\begin{align}
i \partial_z \psi(i,z) = H \psi(i,z)
,
\end{align}
where $\psi(i,z)$ represents the wave function at $i$-th waveguide, $z$ is the propagation axis of the light, and $H$ is the Hamiltonian of the quasiperiodic network models with slowly varying potentials.
For localized states, the output light tends to be localized in one of the waveguides, while for extended states, it tends to occupy the whole chain. By varying system parameters, such as the potential strength and hopping term, one can explore the transitions between localized and extended states. 

In Fig.~\ref{fig-fig9}(b), we plot the fractal dimension of the eigenstates as a function of energy $E$ and potentail strength $V$ for a small system ($L = 100$), where we can also see clear MEs. Here we consider quasiperiodic mosaic model with slowly varying potentials when $\kappa = 2$. To simulate the experimental results, we plot the evolution of the wave function for various potential strength $V$ in Figs.~\ref{fig-fig9}(c) - \ref{fig-fig9}(f). Although the specific evolution of wave function depends strongly on its initial state, the properties of localized state or extended state remain the same. Here, we set the initial state as $\psi(i,0) = \ket{0,\cdots,1_{i=50},\cdots,0}$. For $V = 0.5t$, the wave function gradually spreads through the entire system. As the potential strength increases to $V = t$, the wave function propagating to the boundary decreases, since the number of extended states in the system is reduced. When $V = 2t$ and $V = 3t$, the wave functions become localized around the center of the system, as all states away from the band center are localized, which we leave as an open question to be investigated in the future. More information about the experiment setups and their tight-binding model can be found in Refs. \cite{chen2025reentrant, Chen2021Tight,  Wang2022EdgeState}. In these experiments, the inter-chain coupling and its o-site potential can be designed almost arbitrary. 

Here, we only demonstrate the possible observation of localization-delocalization transition with the increasing of potential strength in network models in some small systems. The precise phase boundary from this kind of measurement can not be precisely determined in the current stage, which is a common problem in all quantum simulating systems \cite{Georgescu2014Quantum, Heras2014Digital, Altman2021Quantum}; see our discussion in the next section. To this end, new algorithm to exact the MEs from the evolution is required in future. 

\section{Conclusions}\label{sec-conclusions}
To conclude, in this work we propose a new class of 1D quasiperiodic network models slowly varying potentials and we can exactly determine their MEs. Unlike the previously proposed quasiperiodic network models \cite{hu2025hidden}, this model doesn't exhibit hidden self-duality, since the slowly varying potential is not self-dual. We integrate out the periodic sites to obtain an effective model, where $g(E)V$ represents the effective potential and $f(E)$ denotes the effective eigenenergy. Substituting $f$ and $g$ into $f = \pm(2t^\kappa \pm gV)$ will give the concrete MEs, following Das Sarma {\it et. al.} \cite{das1988mobility}. We confirm these results using numerical method. With this method, we can study the exact MEs with slowly varying potentials in some much more complicated models, including non-Hermitian ones. \textcolor{black}{Additionally, we also study the connection between topological transition and localization-delocalization transition induced by a complex phase}, as discussed in Refs. \cite{Padhan2024Complete, Loughi2019Topological}. In this way, the exact MEs in models with slowly varying potentials should greatly broaden our understanding of MEs in 1D disordered models. 

Now, it is quite possible that we are in a position to reflect on the possible future of quantum simulations. In this work, we show that the rich phases in the quasiperiodic network models can be simulated using the optical waveguides. In actually, there are much more platforms to simulate these models, including optical and acoustic waveguides \cite{Khelif2004Guiding, Apigo2019Observation, Coutant2021Acoustic}, ultracold atoms, superconducting qubits \cite{You2011Atomic, Jiang2011Interface, Bialczak2011Fast, Jeffrey2014Fast, Wang2014Measurement, Brosco2024Superconducting}, in which most of the essential physics in these models can be obtained using some small quantum systems. In recent years, these small systems have been used to demonstrate these fundamental physics. Yet, the detailed physics, including the scaling laws and the phase boundaries have not been precisely determined. This is different from that in the community of condensed matter physics, in which the phase transition and the associated scaling exponents have been precisely determined in experiments. The similar accuracy has not yet been achieved in quantum simulation. In the future, with the increasing of experimental techniques, it is quite possible that these simulating platforms can be used to determine the precise phase boundary and the associated critical exponents, making quantum simulations a powerful tool to explore possible new physics. At the current stage, only some intriguing phenomena have been simulated. If this kind of accuracy can be achieved, the physical models may serve as an important platform for exploring Anderson transition in the network models \cite{Piotr2023Universality}, which will help us to resolve the long-standing problems in this physics. Especially, in the presence of many-body interaction \cite{Iyer2013Many, Vosk2015Theory, Bere2015Many} and in the higher dimensional models with various potentials \cite{Arguello2020Quamtum, Ott2021Scalable}, these physics cannot be efficiently simulated using classical computers, however, it is quite possible that  these new physics can be simulated using these intriguing platforms. In the past ten years, we have witnessed the huge progress in quantum simulation with these waveguides, from two coupled waveguides to large networks without thousands of waveguides, and thus in the next decade, We will optimistically to see quantum simulation, as a new means of research, to unveil new physical phenomena.

\begin{acknowledgements}
This work is supported by the Strategic Priority Research Program of the Chinese
Academy of Sciences (Grant No. XDB0500000),  the National Natural Science Foundation of China (Grant No. 12374017, No. 12074362 and No. U23A2074) and the Innovation Program for Quantum Science and Technology (2021ZD0303303, 2021ZD0301200, 2021ZD0301500). A.-M. G. is supported by the NSFC (Grant Nos. 12274466 and 11874428) and the Hunan Provincial Science Fund for Distinguished Young Scholars (Grant No. 2023JJ10058).
\end{acknowledgements}

\bibliography{ref.bib}

\end{document}